\pdfoutput=1
\documentclass[aps,prl,twocolumn,showpacs,superscriptaddress,groupedaddress]{revtex4}  
\usepackage{graphicx,epsfig}  
\usepackage{dcolumn}   
\usepackage{bm}        
\usepackage{amssymb,amsmath}   
\usepackage{comment}
\def\a {\alpha}
\def\ve {\varepsilon}
\def\e {\epsilon}
\def\b {\beta}
\def\g {\gamma} \def \G {\Gamma}
\def\d {\delta} 
\def\l {\lambda} \def\L {\Lambda}
\def\m {\mu}
\def\n {\nu}
\def\p {\pi}
\def\r {\rho}
\def\s {\sigma}

\def\dd{{\rm d}}

\def\({\left(}
\def\){\right)}
\def\[{\left[}
\def\]{\right]}
\def\Disc{\mathrm{Disc}}

\def\beq {\begin{equation}}
\def\eq {\end{equation}}
\def\bea{\begin{eqnarray}}
\def\eea{\end{eqnarray}}
\def\bf{\textbf}
\newcommand{\notd}[1] { \setbox0=\hbox{$#1$}
\dimen0=\wd0   \setbox1=\hbox{/} \dimen1=\wd1  \ifdim\dimen0>\dimen1
 \rlap{\hbox to \dimen0{\hfil/\hfil}}  #1 \else \rlap{\hbox to \dimen1{\hfil$#1$\hfil}}  /  \fi  }

\def\dd{{\rm d}}
\def\nn{\nonumber}
\def\a {\alpha}
\def\ve {\varepsilon}
\def\e {\epsilon}
\def\b {\beta}
\def\g {\gamma} \def \G {\Gamma}
\def\d {\delta} 
\def\l {\lambda} \def\L {\Lambda}
\def\m {\mu}
\def\n {\nu}
\def\p {\pi}
\def\r {\rho}
\def\s {\sigma}

\def\dd{{\rm d}}

\def\({\left(}
\def\){\right)}
\def\[{\left[}
\def\]{\right]}
\def\Disc{\mathrm{Disc}}

\def\beq {\begin{equation}}
\def\eq {\end{equation}}
\def\bea{\begin{eqnarray}}
\def\eea{\end{eqnarray}}
\def\bf{\textbf}

\def\dd{{\rm d}}
\def\nn{\nonumber}

\begin{document}

\title{Anomalous magnetic moment of the muon in a dispersive approach}

\author{Vladyslav Pauk}
\author{Marc Vanderhaeghen}
\affiliation{Institut f\"ur Kernphysik, Johannes Gutenberg-Universit\"at,  Mainz, Germany}
\affiliation{
PRISMA Cluster of Excellence, Johannes Gutenberg-Universit\"at,  Mainz, Germany}

\date{\today}

\begin{abstract}
We present a new general dispersive formalism for evaluating the hadronic light-by-light scattering contribution to the anomalous magnetic moment of the muon. In the suggested approach, this correction is related to the imaginary part of the muon's electromagnetic vertex function. The latter may be directly related to measurable hadronic processes by means of unitarity and analyticity. As a test we apply the introduced formalism to the case of meson pole exchanges and find agreement with the direct two-loop calculation.
\end{abstract}

\maketitle

The keen interest in the anomalous magnetic moment of the muon $a_\m$ is motivated by its high potential for probing physics beyond the Standard Model (SM). The presently observed $3-4\s$ discrepancy~\cite{Blum:2013xva} allows for a number of beyond SM scenarios which relate this deviation to contributions of hypothetical particles, see~\cite{Queiroz:2014zfa} and references therein. On the experimental side, the new measurements both at Fermilab (E989) \cite{LeeRoberts:2011zz} as well as at J-PARC ~\cite{Iinuma:2011zz} aim to reduce the experimental error on $a_\m$ to $\d a_\m(\rm exp)=\pm 16\times 10^{-11}$, which is a factor of four improvement over the present value. The expected precision of the new experiments will give access to scales up to  $\L\sim m/\sqrt{\d a_\m}\sim8$ TeV, where $m$ is the mass of the muon \cite{Czarnecki:2001pv}, which makes it highly competitive to measurements at the Large Hadron Collider (LHC). However, the interpretation of $a_\m$ is undermined by theoretical uncertainties of the strong-interaction contributions entering its SM value. Depending on the analysis of these hadronic contributions~\cite{Blum:2013xva, Jegerlehner:2013sja} the present SM uncertainty amounts to the range $\d a_\m(SM)=\pm(49 - 58) \times 10^{-11}$ which significantly exceeds the future experimental accuracy. This motivates an intense activity to reliably estimate contributions of hadrons to $a_\m$, see \cite{Benayoun:2014tra} and references therein. 

\begin{figure}[h]
\centering
\vspace{.4cm}
\includegraphics[width=3cm]{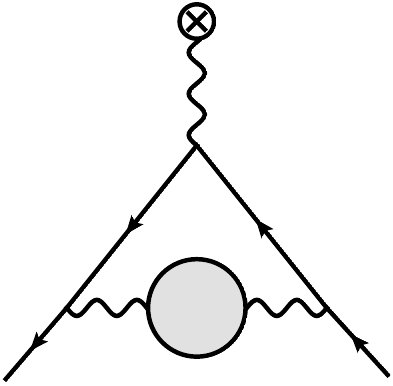}\hspace{.8cm}
  \includegraphics[width=3.4cm]{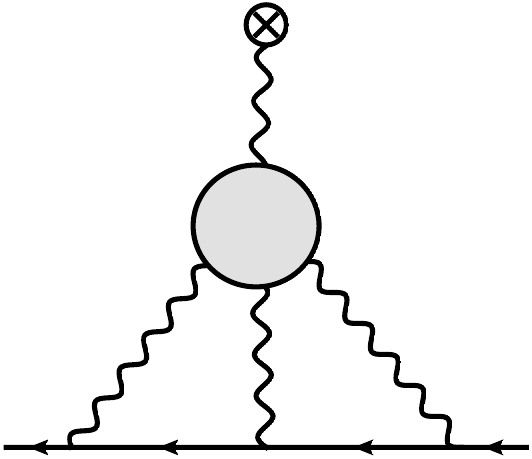}\\
  \label{triangle}
  \caption{\label{fig:LbL}The hadronic vacuum polarization (left panel) and light-by-light scattering (right panel) contributions to the anomalous magnetic moment of the muon}
\end{figure}

The hadronic uncertainties mainly originate from hadronic vacuum polarization (HVP) and hadronic light-by-light (HLbL) insertion diagrams shown in Fig. \ref{fig:LbL}. The dominant HVP contribution can be reliably estimated on the basis of experimental information of electromagnetic hadron production processes implemented via the dispersion technique.
The existing estimates are based on data for $e^+ e^- \to \mathrm{hadrons}$, data for $e^+ e^- \to \gamma + \mathrm{hadrons}$, as well as $\tau$ decays~(see~\cite{Blum:2013xva} and references therein) yielding an accuracy $\delta a_\mu (\mathrm{l.o. \, HVP}) = \pm42.4 \times 10^{-11}$ \cite{Jegerlehner:2013sja}. The ongoing experiments at $e^+e^-$-colliders (mainly VEPP-2000 and BES-III) 
will provide valuable experimental input to further 
constrain this contribution. It was estimated in \cite{Blum:2013xva} that 
the forthcoming data will allow to reduce the uncertainty in the HVP by around a factor of two.

Unlike the HVP contribution, in most of the existing estimates of the HLbL contribution, the description of the non-perturbative light-by-light matrix element is based on hadronic models rather than determined from data. These approximations are based on a requirement of consistency with the asymptotic constraints of QCD and predict that the hadronic corrections are dominated by long-distance physics, namely due to exchange of the lightest pseudoscalar states \cite{Jegerlehner:2009ry}. Unfortunately, a reliable estimate based on such models is possible only within certain kinematic regimes. This results in a large, mostly uncontrolled uncertainty of $a_\m$. The two main estimates of the HLbL contribution to $a_\m$ yield:
\begin{eqnarray}
 a_\mu (\mathrm{HLbL}) &=& ( 116 \pm 39) \times 10^{-11} \quad \quad 
\mathrm{Ref.}~\mbox{\cite{Jegerlehner:2009ry}}, 
\label{eq:hlbl} \\
 a_\mu (\mathrm{HLbL}) &=& ( 105 \pm 26) \times 10^{-11} \quad \quad 
 \mathrm{Ref.}~\mbox{\cite{Prades:2009tw}}.
 \end{eqnarray}
To overcome the model dependence one may resort to data-driven approaches for the HLbL contribution to $a_\mu$.
Recently, such an approach based on the analytic structure of the HLbL tensor has been discussed in \cite{Colangelo:2014dfa,Colangelo:2014pva}. In the present work, we present a new data driven approach for calculating $a_\m$ based on the analytic properties of the muon's electromagnetic vertex function. We express $a_\m$ through a dispersive integral over the discontinuity of the muon's electromagnetic vertex function, which in turn can be related to observables.

Defined as a static limit ($k^2=0$, with $k$ being the photon momentum) of the Pauli form factor $F_2(k^2)$, the anomalous magnetic moment can be extracted from the vertex function by a projection technique as was elaborated in~\cite{Barbieri:1972as, Roskies:1990ki}. Applying the Feynman rules to the diagram on the right panel of Fig.~\ref{fig:LbL} and rewriting the virtual photon propagators using the completeness relation for photon polarization vectors as $g_{\m\n}/(q^2-i\ve)=\sum\limits_{\l}(-1)^\l \ve_\m(q,\l)\ve_\n^\ast(q,\l)/(q^2-i\ve)$, the HLbL contribution to $F_2(k^2)$ is obtained as a two-loop integral

\begin{widetext}
\begin{eqnarray}
F_2(k^2)=e^6\sum\limits_{\l_1,\l_2,\l_3,\l}(-1)^{\l+\l_1+\l_2+\l_3}&&\hspace{-.5cm}\int\frac{\dd^4 q_1}{(2\p)^4}\int\frac{\dd^4 q_2}{(2\p)^4}L_{\l_1\l_2\l_3\l}(p,q_1,k-q_1-q_2,q_2)\nn\\
\hspace{4.5cm}&\times&\frac{\Pi_{\l_1\l_2\l_3\l}(q_1,k-q_1-q_2,q_2,k)}{q_1^2q_2^2(k-q_1-q_2)^2\[(p+q_1)^2-m^2\]\[(p+k-q_2)^2-m^2\]}.\label{eq:F2loopexplicit}
\end{eqnarray}
In Eq.~(\ref{eq:F2loopexplicit}) the fourth-rank hadronic vacuum polarization tensor projected on the helicity basis is defined as a Fourier transform of the four-current correlator in the QCD vacuum $\left|\Omega\right>$:
\begin{eqnarray}
\Pi_{\l_1\l_2\l_3\l_4}(q_1,q_2,q_3)&=&\e^\m(q_1,\l_1)\e^\n(q_2,\l_2)\e^\l(q_3,\l_3)\e^\r(q_4,\l_4)\label{eq:4pointLbL}\\
&\times&\int\dd^4x_1\int\dd^4x_2\int\dd^4x_3\,e^{i(q_1\cdot x_1+q_2\cdot x_2+q_3\cdot x_3)}\left<\Omega\left|\mathrm{T}\{j_\m(x_1)j_\n(x_2)j_\l(x_3)j_\r(0)\}\right|\Omega\right>.\nn
\end{eqnarray}
The leptonic coefficient functions $L_{\l_1\l_2\l_3\l_4}$ are defined by
\begin{eqnarray}
L_{\l_1\l_2\l_3\l_4}(p,q_1,q_2,q_3)&=&\ve^\ast_{\m}(\l_1,q_1)\ve^\ast_{\n}(\l_2,q_2)\ve^\ast_{\l}(\l_3,q_3)\ve^\ast_{\s}(\l_4,q_4)\\
&\times&\mathrm{Tr}\[\L^\s(p+q_1+q_2,p)\g^\l(\notd{p}+\notd{q}_1+\notd{q}_2+m)\g^\n(\notd{p}+\notd{q}_1+m)\g^\m\],\nn
\end{eqnarray}
with projector
\begin{eqnarray}
\L_\s(p',p)&=&\frac{m^2}{k^2(4m^2-k^2)}(\notd{p}+m)\[\g_\s+\frac{k^2+2m^2}{m(k^2-4m^2)}(p'+p)_\s\](\notd{p}'+m).\nn
\end{eqnarray}
\end{widetext}
In the latter formulae $p'$ and $p$ denote momenta of the muon before and after scattering on an electromagnetic field with $k=p'-p$, $q_i$ and $\l_i$ are the virtual photons' momenta and helicities.

When analytically continued to complex values of the external photon's virtuality $k^2$, the muon's electromagnetic vertex function possesses branch point singularities joining the physical production thresholds, as is dictated by unitarity \footnote{The anomalous thresholds for the three-point functions located below the normal thresholds do not appear in the considered case.}. Using Cauchy's integral theorem, the form factor in Eq.~(\ref{eq:F2loopexplicit}) can be represented as an integral along a closed contour avoiding the cuts and extended to infinity. Assuming that the form factor vanishes uniformly when $k^2$ tends to infinity the contour integral reduces to an integral of the form factor's discontinuity $\Disc_{k^2}F_2(k^2)$ along the cut in the $k^2$-plane starting from the lowest branch point:
\beq
F_2(0)=\frac1{2\p i}\int\limits_0^\infty\frac{\dd k^2}{k^2}\mathrm{Disc}_{k^2}\,F_2(k^2).
\label{eq:F2disp}
\eq

As can be seen from the structure of the two-loop integral in Eq.~(\ref{eq:F2loopexplicit}), the branch cuts of the Pauli form factor $F_2(k^2)$ are related to the propagators of virtual particles and non-analyticities of the HLbL tensor. 
The latter possesses two types of discontinuities, the corner (one-photon) and cross (two-photon) cuts. The corner cuts are related to a conversion of a photon to a hadronic state with negative $C$-parity, while the cross cuts are related to a two-photon production of a $C$-even hadronic state. As the dominant contributions originate from the lowest thresholds it is mainly governed by intermediate states including pions. In particular, the lowest threshold in the $C$-odd channel is related to a $\p^+\p^-$-pair production and in the $C$-even channel to a $\p^0$ intermediate state. By virtue of unitarity, these discontinuities are related to amplitudes of physical hadron production processes. Experimentally, the amplitudes involved in the unitarity equation for the required discontinuities can be measured in two-photon and $e^+e^-$ production processes (for references see \cite{Benayoun:2014tra}). The dispersive analysis of the two-pion production channel by a real and a virtual photon was recently discussed in~\cite{Moussallam:2013una}.

\begin{figure}[h]
\centering
  \includegraphics[width=7.8cm]{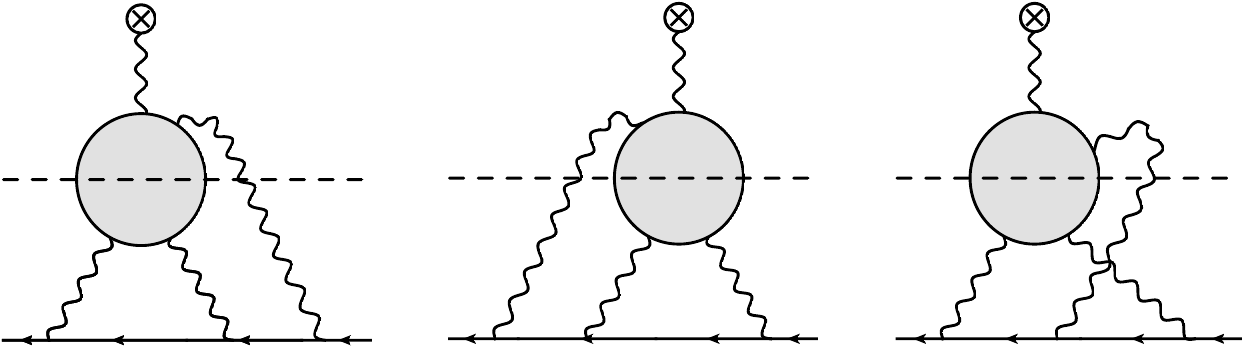}
  \vspace{.3cm}\\
  \includegraphics[width=7.8cm]{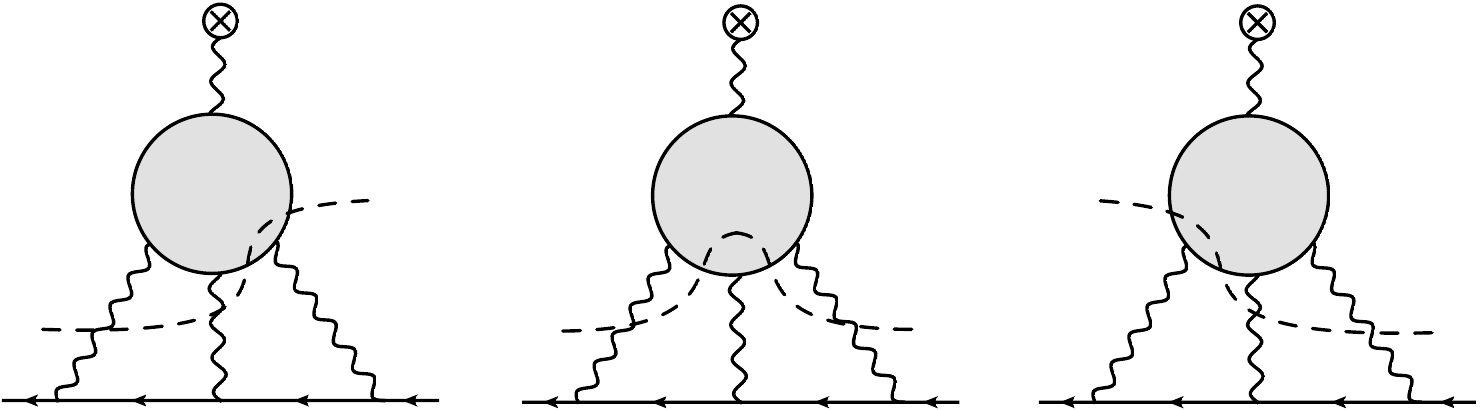}
  \vspace{.3cm}\\
  \includegraphics[width=2.5cm]{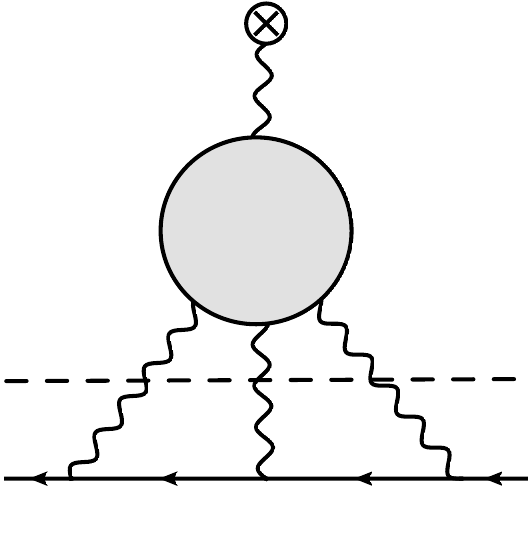}
  \hspace{.2cm}
    \includegraphics[width=2.5cm]{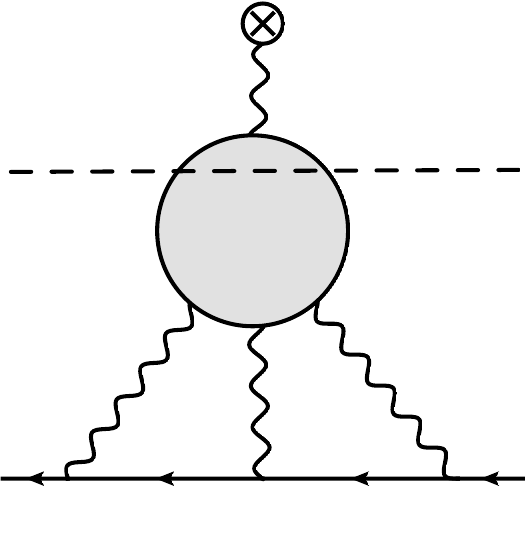}
  \hspace{.2cm}
  \includegraphics[width=2.5cm]{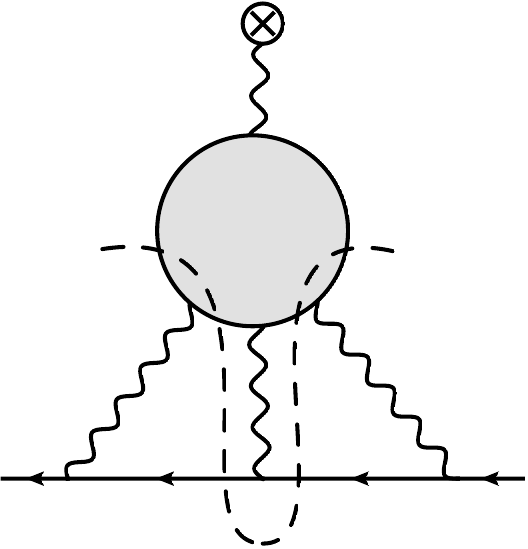}
  \caption{\label{fig:undiagr}Unitarity diagrams contributing to the imaginary part of the vertex function. The cut indicates the on-shell intermediate state. }
\end{figure}

Taking into account the analytical structure of the HLbL tensor, the discontinuity in Eq.~(\ref{eq:F2disp}) is obtained as a sum of nine topologically different contributions, which are graphically represented by unitarity diagrams in Fig.~\ref{fig:undiagr}. On a practical level, the contribution of a particular unitarity diagram is obtained by replacing the cut virtual propagators in the two-loop integral by corresponding delta functions, and the cut vertices by their appropriate discontinuities. As an example for the first diagram in Fig.~\ref{fig:undiagr}, it implies

\begin{widetext}
\begin{eqnarray}
\Disc_{}F_2(k^2)&=&e^6\sum\limits_{\l_1,\l_2,\l_3,\l}(-1)^{\l+\l_1+\l_2+\l_3}\int\frac{\dd^4 q_1}{(2\p)^4}\int\frac{\dd^4 q_2}{(2\p)^4}\frac1{q_1^2}\frac1{(k-q_1-q_2)^2}\frac1{(p+q_1)^2-m^2}\frac1{(p+k-q_2)^2-m^2}\nn\\
&\times&L_{\l_1\l_2\l_3\l}(p,q_1,k-q_1-q_2,q_2)(2\p i)\d(q_2^2)\Disc_{(k-q_2)^2}\Pi_{\l_1\l_2\l_3\l}(q_1,k-q_1-q_2,q_2,k).
\label{eq:discpi}
\end{eqnarray}
\end{widetext}
The non-perturbative discontinuity function $\Disc_{(k-q_2)^2}\Pi_{\l_1\l_2\l_3\l}$ in Eq.~(\ref{eq:discpi}) is directly related to amplitudes of processes $\g^\ast\g^\ast\to X$ and $\g^\ast\to\g X$, with $X$ denoting a $C$-even hadronic state, which are accessible experimentally.

\begin{figure}[h]
\centering
  \includegraphics[width=8.6cm]{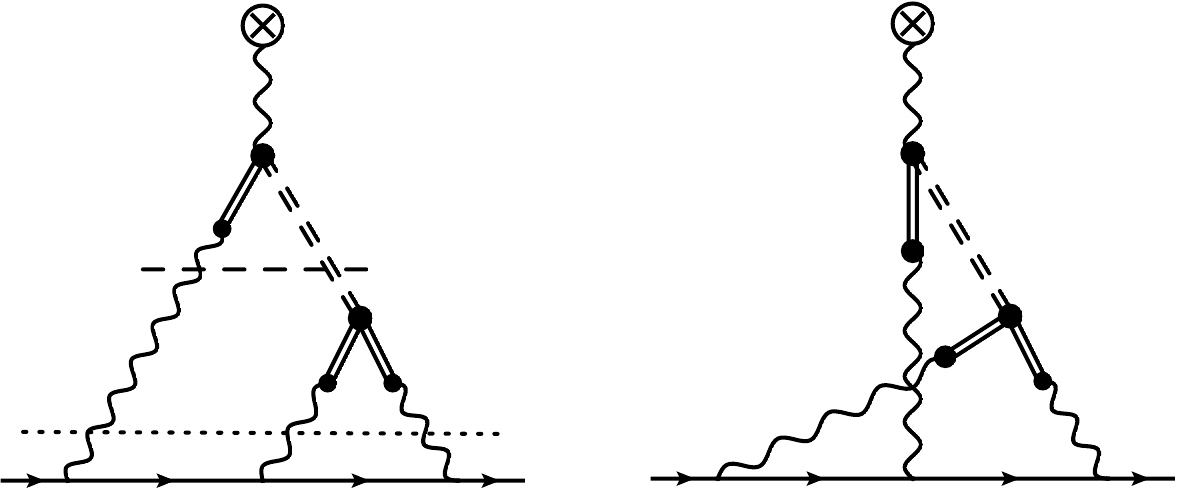}
  \caption{\label{fig:LbLpole}The two topologies of the HLbL contribution to $a_\m$ in the pole approximation and examples of the two-particle (dashed) and three-particle (dotted line) cuts for the first topology (left panel). The wavy lines stand for photons, whereas the double-dashed (double-solid) lines stand for pseudoscalar (vector) meson poles.}
\end{figure}

To set up and test the technique for evaluating the phase space and dispersion integrals we consider a well-studied approximation for the contribution of a pseudoscalar meson (corresponding to $\p^0$, $\eta$ and $\eta'$ exchanges), based on the large-$N_c$ limit~\cite{Knecht:2001qf}. In such approximation the HLbL amplitude is approximated by a pole term of the form:
\begin{align}
\Pi&_{\rm pole}(q_1^2,(k-q_1-q_2)^2,q_2^2,k^2,(k-q_1)^2,(q_1+q_2)^2)\\
&=\frac{|F(0,0,M^2)|^2}{(q_1^2-\L^2)(q_2^2-\L^2)((k-q_1-q_2)^2-\L^2)(k^2-\L^2)}\nn\\
&+\mathrm{crossed\; terms}\nn.
\end{align}
Here $M$ and $\L$ denote masses of the pseudoscalar and vector mesons respectively and $F(0,0,M^2)$ stands for the pseudoscalar meson transition strength into real photons.
While the analytical structure of the HLbL amplitude in the $C$-even channel is defined by a pole due to an exchange of the pseudoscalar meson, in the $C$-odd channel it is governed by a vector state exchange which can be confronted with the Vector Meson Dominance model (see \cite{Czerwinski:2012ry} for a review). The analytical structure of the two distinct contributions to the muon's electromagnetic vertex function arising from such pole terms is equivalent to the structure of the two-loop diagrams shown in Fig.~\ref{fig:LbLpole}.

We demonstrate the process of computation on the example of the first topology illustrated by a diagram in the left panel of Fig.~\ref{fig:LbLpole}. The contribution of the second topology has a similar structure and is computed in an analogous way. We can consider the dispersive integral for $F_2(k^2)$ multiplied by $(k^2-\L^2)$, which removes the pole in $k^2$ and its related discontinuity. The remaining discontinuities may be separated in two and three-particle cuts. The two-particle cuts include the $\g \p^0$ and $\r \p^0$ intermediate states. The three-particle cuts include: $\g\g\g$, $\g\g\r$, $\g\r\g$, $\r\g\g$, $\g\r\r$, $\r\g\r$, $\r\r\g$, $\r\r\r$ intermediate states. Graphically they are represented by cuts shown in the left panel of Fig.~\ref{fig:LbLpole} for the case of $\g \p^0$ (two-particle) and $\g\g\g$ (three-particle) intermediate states. 

Following the procedure described above, we replace the propagators of the cut photons and mesons by the corresponding on-shell delta-functions.  
Thus for instance the $\g \p^0$ cut is obtained by
\begin{widetext}
\begin{eqnarray}
\Disc_{\p^0\g}F_2(k^2)&=&e^6\L^6F_{P\g^\ast\g^\ast}(0,0,M^2)\int\frac{\dd^4 q_1}{(2\p)^4}\int\frac{\dd^4 q_2}{(2\p)^4}(2\p i)^2\d((k-q_1)^2-M^2)\d(q_1^2-\L^2)\frac1{q_1^2} \frac1{q_2^2-\L^2}\\
&\times&\frac1{q_2^2}\frac1{(k-q_1-q_2)^2}\frac1{(k-q_1-q_2)^2-\L^2}\frac1{(p+q_1)^2-m^2}\frac1{(p+k-q_2)^2-m^2}L(p,q_1,k-q_1-q_2,q_2)\nn,
\end{eqnarray}
where 
\begin{eqnarray}
L(p,q_1,q_2,q_3)&=&\mathrm{Tr}\[\L^\s(p+q_1+q_2,p)\g^\l(\notd{p}+\notd{q}_1+\notd{q}_2+m)\g^\n(\notd{p}+\notd{q}_1+m)\g^\m\]\\&\times&\ve_{\m\s\a\b} q_1^\a (q_1+q_2+q_3)^\b \ve_{\n\l\g\d} q_2^\g q_3^\d\nn.
\end{eqnarray}
\end{widetext}
The phase-space integrals and the one-loop insertions are evaluated partially analytically with the subsequent numerical computation, see~\cite{Pauk:2014jza} for some technical details in the case of scalar field theory. More details of the present calculation will be given elsewhere. 
The lowest threshold for the two-particle cut is located at $k^2=M^2$ corresponding to $\g\p^0$ intermediate state. For the three-particle discontinuity it is $k^2=0$ related to the $\g\g\g$ cut. Thus the dispersion integral has the form
\beq
F_2(0)=\frac1{2\p i}\int\limits_{M^2}^{\infty}\frac{\dd k^2}{k^2}\Disc_2 F_2(k^2)+\frac1{2\p i}\int\limits_{0}^\infty\frac{\dd k^2}{k^2}\Disc_3 F_2(k^2)
\eq
with $\Disc_2 F_2(k^2)$ and $\Disc_3 F_2(k^2)$ denoting the sum of two- and three-particle discontinuities. 

\begin{figure}[h]
\centering
  \includegraphics[width=8.5cm]{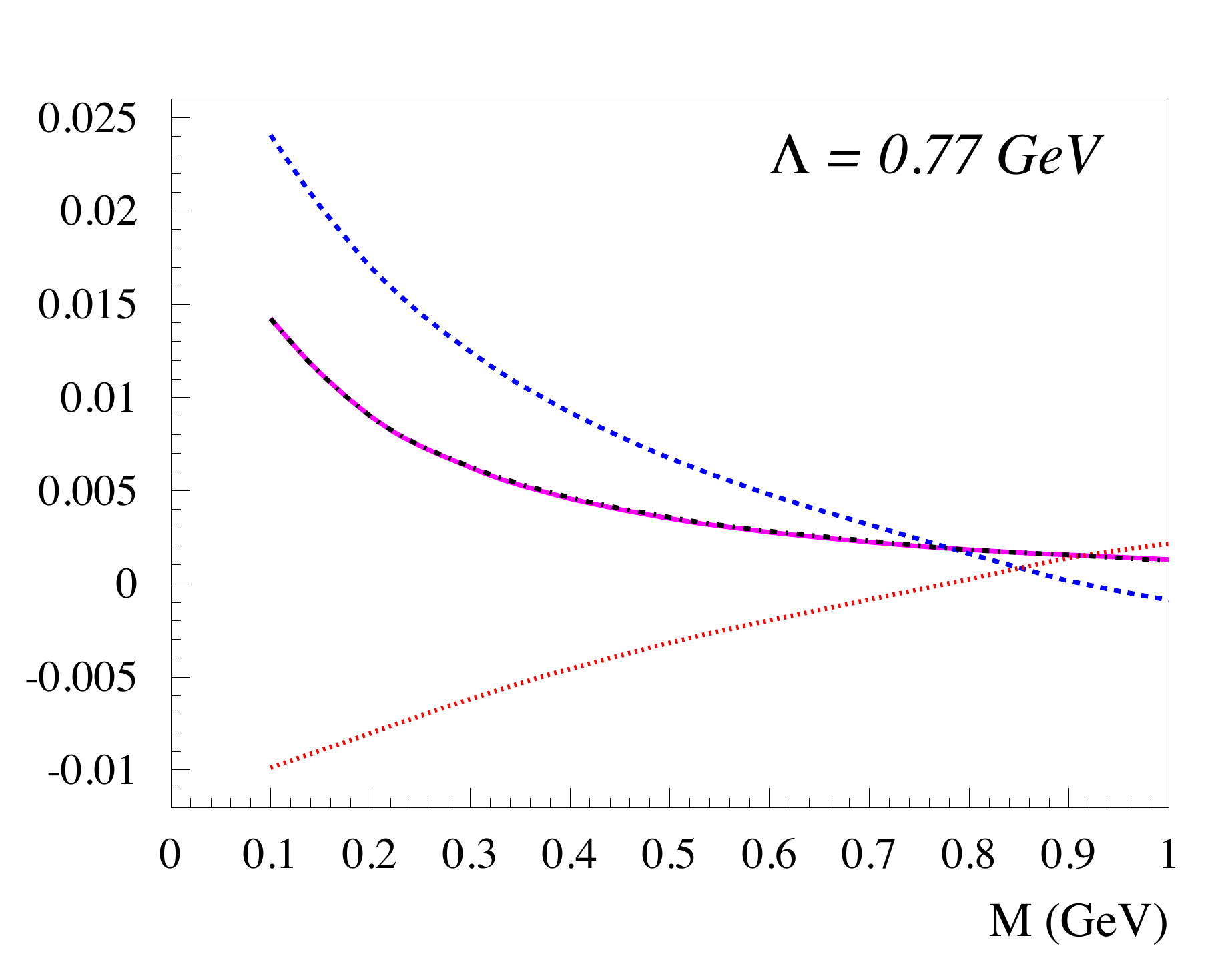}
  \caption{The value of the HLbL pole contribution due to the diagram of topology (1) (left panel in Fig.~\ref{fig:LbLpole}) to $a_\m$ scaled by factor of $4\p M^3/(e^2 \G_{\g\g})$ depending on the mass of the pseudoscalar meson, with $\G_{\g\g}$ the two-photon decay width of the pseudoscalar meson. The blue dashed (red dotted) curve represents the contribution of the two (three) particle cuts. Their sum is denoted by the black dashed-dotted curve. The result of the direct evaluation of the two-loop integral is illustrated by the pink solid curve.}
  \label{fig:impartscalar}
\end{figure}

For a test we analyze the dependence of the HLbL contribution to $a_\m$ on the pseudooscalar meson mass $M$ and compare our result with the calculation using the approach of \cite{Knecht:2001qf}, by evaluating the two-loop integral in Euclidian space.
The contributions of the two types of discontinuities, their sum and the result of the conventional integration depending on the pseudoscalar meson mass are shown in Fig.~\ref{fig:impartscalar}, and their numerical values at the $\p^0$ mass are summarized in Tab.~\ref{tab:table2}. When comparing the result obtained by the two different methods we find an exact agreement. 
\begin{table}
\caption{\label{tab:table2}The contributions to $a_\m$ (in units $10^{-10}$) of two-particle (2p) and three-particle (3p) cuts for the two topologies (see Fig.~\ref{fig:LbLpole}) appearing in the pole approximation compared to the results of the conventional 2-loop integration of \cite{Knecht:2001qf}. Note that total$=2\times(1)+(2)$.}
\begin{ruledtabular}
\begin{tabular}{ccccc}
 &2p-cut& 3p-cut&total&direct\\
\hline
(1)  &  4.91   &  -2.14 & 2.77 & 2.77 \\
(2)  &  -7.40 &  7.56  & 0.16  & 0.16  \\
total& 2.42   &  3.28   & 5.70 & 5.70  \\
\end{tabular}
\end{ruledtabular}
\end{table}

The suggested approach opens a new alternative strategy for evaluating the HLbL contributions to the anomalous magnetic moment of the muon. It implies the dispersive evaluation of the loop integrals. In contrast to the conventional approach where the integration is carried out after the analytical continuation to the Euclidian region, it allows for a more straightforward relation to observables. The fact that the involved matrix elements are partially on-shell simplifies implementation of the experimental data which presently is not available for totally off-shell matrix elements. Practically, computations within this approach involve both analytical and numerical evaluations of phase-space and dispersion integrals. To test the numerical algorithms we considered the well studied model of the pole contribution. The dispersive evaluation shows good numerical stability and exact agreement with the existing result. Further development of the dispersion technique and its data driven applications give promising perspectives for a reduction of the hadronic uncertainties. 
In particular an important realistic application of the suggested method concerns the contribution due to the two-pion intermediate states which is presently the largest source of uncertainty in HLbL correction to $a_\m$. 

This work was supported by the Deutsche Forschungsgemeinschaft DFG in part through the Collaborative 
Research Center ``The Low-Energy Frontier of the Standard Model" (SFB 1044), and in part through 
the Cluster of Excellence "Precision Physics, Fundamental Interactions and Structure of Matter" (PRISMA).


\begin{thebibliography}{99}

\bibitem{Blum:2013xva} 
  T.~Blum, A.~Denig, I.~Logashenko, E.~de Rafael, B.~Lee Roberts, T.~Teubner and G.~Venanzoni,
  arXiv:1311.2198 [hep-ph].

\bibitem{Queiroz:2014zfa} 
  F.~S.~Queiroz and W.~Shepherd,
  Phys.\ Rev.\ D {\bf 89}, 095024 (2014)
  [arXiv:1403.2309 [hep-ph]].


  \bibitem{LeeRoberts:2011zz} 
  B.~Lee Roberts [Fermilab P989 Collaboration],
  Nucl.\ Phys.\ Proc.\ Suppl.\  {\bf {218}}, 237 (2011).
  
\bibitem{Iinuma:2011zz} 
  H.~Iinuma [J-PARC New g-2/EDM experiment Collaboration],
  J.\ Phys.\ Conf.\ Ser.\  {\bf {295}}, 012032 (2011).

\bibitem{Czarnecki:2001pv} 
  A.~Czarnecki and W.~J.~Marciano,
  Phys.\ Rev.\ D {\bf 64}, 013014 (2001)
  [hep-ph/0102122].
 
   \bibitem{Jegerlehner:2013sja} 
  F.~Jegerlehner,
  Acta Phys.\ Polon.\ B {\bf {44}}, no. 11, 2257 (2013).

\bibitem{Benayoun:2014tra} 
  M.~Benayoun, J.~Bijnens, T.~Blum, I.~Caprini, G.~Colangelo, H.~Czy$\rm\acute{z}$, A.~Denig and C.~A.~Dominguez {\it et al.},
  arXiv:1407.4021 [hep-ph].

 \bibitem{Jegerlehner:2009ry} 
  F.~Jegerlehner and A.~Nyffeler,
  Phys.\ Rept.\  {\bf 477}, 1 (2009); 
  F.~Jegerlehner,
  Springer Tracts Mod.\ Phys.\  {\bf 226} (2008) 1.
 
\bibitem{Prades:2009tw} 
  J.~Prades, E.~de Rafael and A.~Vainshtein,
  (Advanced series on directions in high energy physics. 20)
  [arXiv:0901.0306 [hep-ph]].
  
\bibitem{Colangelo:2014dfa} 
  G.~Colangelo, M.~Hoferichter, M.~Procura and P.~Stoffer, 
  arXiv:1402.7081 [hep-ph].

\bibitem{Colangelo:2014pva} 
  G.~Colangelo, M.~Hoferichter, B.~Kubis, M.~Procura and P.~Stoffer,
  arXiv:1408.2517 [hep-ph].
 
  
\bibitem{Barbieri:1972as} 
  R.~Barbieri, J.~A.~Mignaco and E.~Remiddi,
  Nuovo Cim.\ A {\bf 11}, 824 (1972).
  
\bibitem{Roskies:1990ki} 
  R.~Z.~Roskies, M.~J.~Levine and E.~Remiddi,
  Adv.\ Ser.\ Direct.\ High Energy Phys.\  {\bf 7}, 162 (1990).

\bibitem{Moussallam:2013una} 
  B.~Moussallam,
  arXiv:1305.3143 [hep-ph].
  
 
\bibitem{Knecht:2001qf} 
  M.~Knecht and A.~Nyffeler,
  Phys.\ Rev.\ D {\bf 65}, 073034 (2002)
  [hep-ph/0111058].

\bibitem{Czerwinski:2012ry} 
  E.~Czerwinski, S.~Eidelman, C.~Hanhart, B.~Kubis, A.~Kupsc, S.~Leupold, P.~Moskal and S.~Schadmand,
  arXiv:1207.6556 [hep-ph].


\bibitem{Pauk:2014jza} 
  V.~Pauk and M.~Vanderhaeghen,
  arXiv:1403.7503 [hep-ph].
  
\end{thebibliography}
\end{document}